\documentclass[10pt,twocolumn,showpacs,preprintnumbers,amsmath,amssymb,aps,prb,longbibliography,superscriptaddress]{revtex4-1}
\usepackage{amsmath}
\usepackage{graphicx}
\usepackage{wasysym}
\usepackage{amsfonts}
\usepackage{amssymb}
\usepackage{bm}
\usepackage{enumerate}
\usepackage{color}
\usepackage{hyperref}

\usepackage{times}
\newcommand{\beq}{\begin{equation}}
\newcommand{\eeq}{\end{equation}}
\newcommand{\bea}{\begin{aligned}}
\newcommand{\eea}{\end{aligned}}
\newcommand{\bes}{\begin{split}}
\newcommand{\ees}{\end{split}}

\newcommand{\RNum}[1]{\uppercase\expandafter{\romannumeral #1\relax}}

\mathchardef\nss="711B

\def\nss{\mathcal{S}}

\newcommand{\normord}[1]{:\mathrel{#1}:}

\def\be{\begin{eqnarray}}
\def\ee{\end{eqnarray}}

\newlength{\myL}

\begin{document}

\title{Interacting 1D Chiral Fermions with Pairing: Transition from Integrable to Chaotic}
\author{Biao Lian}
\affiliation{Department of Physics, Princeton University, Princeton, New Jersey 08544, USA}

\begin{abstract}
We study a generic one-dimensonal quantum model of two flavors (pseudospins) chiral complex fermions by exact diagonalization, which can have local interflavor interaction and superconducting pairings (with all irrelevant terms ignored). Analytically, the model has two solvable (integrable) points in the parameter space: it is a free fermion model when the fermion interaction is zero, and is a free boson Luttinger liquid when there is a global U(1)$^{(\uparrow)}\times$U(1)$^{(\downarrow)}$ symmetry (with nonzero interaction). When the global symmetry of the interacting model is lowered by turning on symmetry breaking parameters, the model undergoes a transition from a quantum integrable model to a fully quantum chaotic model, as we demonstrate by examining the level spacing statistics (LSS) of the many-body energy spectrum. In particular, there is a possibly integrable regime with intermediate global symmetries, where the model is neither free bosons nor free fermions, but shows Poisson LSS in each global symmetry charge sector. This implies the existence of hidden (quasi)local conserved quantities. When the global symmetries are further lowered, the LSS in each charge sector becomes Wigner-Dyson, implying quantum chaos.
\end{abstract}
\maketitle

\section{Introduction}

The study of the integrability of one-dimensional (1D) interacting quantum models has a long history in condensed matter physics. The earliest such exact solution studies date back to the Bethe ansatz for the 1D Heisenberg model \cite{bethe1931} and the Onsager solution for the 2D classical Ising model (which is equivalent to a 1D quantum Ising model) \cite{onsager1944}. The successive studies in the later decades have revealed many more integrable 1D quantum models, such as the Lieb-Liniger model of 1D Bose gas \cite{lieb1963a,lieb1963b}, 1D Hubbard model \cite{lieb1968}, spin models obeying the Yang-Baxter equation \cite{mcguire1964,yang1967,baxter1971}, and the Luttinger liquid of interacting fermions \cite{Dzyaloshinski1996,Haldane_1981,Haldane1981,Tomonaga1950,Luttinger1963,Wen1,Wen2}. In particular, Chen-Ning Yang has made significant contributions to the understanding of thermodynamic behaviors of the integrable spin models \cite{yang1952,yang1966a,yang1966b,yang1966c,yang1966d}, interacting gases \cite{yang1969} in 1D, and the Yang-Baxter equation named partly after him as a sufficient condition for integrability \cite{yang1967}. These developments have significantly advanced the physicists' understanding of quantum integrability, phase transitions and non-equilibrium quantum dynamics \cite{rigol2007,rigol2008} in 1D systems. 

On the other hand, the study of many-body quantum chaos has recently attracted extensive interests. In contrast to quantum integrable models which have enormous number of local or quasilocal conserved quantities \cite{tetelman1981,grabowski1995,ilievski2015,nozawa2020,lian2022conserv}, many-body quantum chaotic systems are expected to have limited number of local conserved quantities (from global symmetries, etc). A class of extensively studied quantum chaotic models is the Sachdev-Ye-Kitaev (SYK) type models \cite{sachdev1992fk,polchinski2016xgd,maldacena2016hyu,kitaev2017awl,lian2019,hu2021chiral}, which are exactly solvable in the large $N$ limit, where $N$ is usually the number of flavors of particles in the model. An indication of the quantum chaos in the SYK models is the positive Lyapunov exponent in the out-of-time-ordered correlation (OTOC) in the large $N$ limit, which has a quantum upper bound $2\pi/\beta$ at temperature $\beta^{-1}$ \cite{Maldacena:2015waa}. Moreover, quantum chaotic systems are expected to show Wigner-Dyson level spacing statistics (LSS) in each symmetry sector of the many-body energy spectrum \cite{Bohigas1984,Dyson1970,Wigner1967}, and usually satisfy the eigenstate thermalization hypothesis (ETH) \cite{jensen1985,deutsch1991,srednicki1994,dalessio2016}. In contrast, integrable systems generically show Poisson LSS \cite{Berry1977}, and violates the ETH.

The scope of this paper is to investigate the integrability and chaos of 1D chiral quantum models, by examining a simplest physical example of chiral fermions. 1D chiral systems constitute a significant class of 1D quantum models, which cannot exist in 1D materials, but can arise as the edge states of 2D gapped chiral topological phases of matter, such as the fractional quantum Hall (FQH) states. In the absence of spatial disorders, a big portion of such chiral models are described by the free boson chiral Luttinger liquid theory \cite{Wen1,Wen2} or free chiral Majorana fermions \cite{moore1991,wen1991}. In 1D chiral models which are not purely chiral (namely, having inequivalent modes propagating in both directions), the symmetry allowed interactions may lead to mode reconstructions under renormalization \cite{haldane1995,Kane1994,Levin2007,Lee2007,levin2013,wangjuven2015,Lian2018,Chamon1994,Wan2002,Wan2003,Sabo2017,Cano2014}, altering the low energy physics. For purely chiral 1D models, the recent studies have revealed a different type of mode reconstruction: the interaction may drive a transition from integrable regimes with well-defined quasiparticles to quantum chaotic regimes without low-energy quasiparticle \cite{lian2019,hu2021chiral,hu2021integrability}. In particular, purely chiral models can have exactly marginal interactions which do not flow under renormalization group, making the physics independent of energy scale. 

A prototypical example is the chiral SYK model of $N$ flavors of chiral Majorana fermions $\psi_i$ ($1\le i\le N$), which has an action \cite{lian2019}
\begin{equation}
\begin{split}
&S_{cSYK}=\int dt dx \mathcal{L}(t,x)\ , \\
&\mathcal{L}=\frac{i}{2}\sum_{i=1}^N \psi_i(\partial_t+\partial_x)\psi_i +\sum_{i<j<k<l}J_{ijkl}\psi_i\psi_j\psi_k\psi_l\ .
\end{split}
\end{equation}
The interactions $J_{ijkl}$ can be taken arbitrarily, and are exactly marginal. For $N\le6$, it is shown that the model with any interactions $J_{ijkl}$ can be exactly solved as a free boson chiral Luttinger liquid (by choosing a proper Majorana fermion basis), thus is integrable. For $N\ge 7$, it is conjectured that the model becomes quantum chaotic and has no quasiparticles. In the large $N$ limit, the quantum chaos can be shown explicitly analytically by $1/N$ expansion techniques \cite{polchinski2016xgd,maldacena2016hyu,kitaev2017awl,lian2019}: by assuming $J_{ijkl}$ are randomly uncorrelated and $\langle J_{ijkl}^2\rangle =\frac{3!J^2}{N^3}$, the velocity-dependent Lyapunov exponent $\lambda_v$ of the OTOC is positive along all velocities within the chiral causality cone of the model, and approaches the maximal chaos bound $2\pi/\beta$ when $J$ approaches the upper-bound $2\pi$ for preserving the chirality (ground state stability).

Such transitions between integrable and chaotic regimes can also arise in 1D chiral models supporting anyons, for instance, in $N$ copies of Wess-Zumino-Witten (WZW) theories with current-current interactions \cite{hu2021chiral}. Moreover, some chiral models can also exhibit properties between the free integrable cases (free bosons or free fermions) and the fully quantum chaotic cases, such as possibly integrable LSS behaviors \cite{hu2021integrability}, and quantum scars \cite{schindler2021exact,martin2021scar}. For instance, the interacting chiral edge states of the $\nu=4/3$ FQH state is recently numerically found to have Poisson LSS in each conserved global symmetry charge sector \cite{hu2021integrability}, indicating the existence of hidden (quasi)local conserved charges and the possibility that the model is integrable. Physically, the low-energy integrability of the chiral edge states is relevant in the detection of their quantum coherent interferences \cite{Willett2009,Zhao2020,Lian2016,Roulleau2008}, for instance, in the Fabry-P\'{e}rot interferometer experiment of the $\nu=1/3$ FQH state \cite{Laughlin1983,Nakamura2020,Carrega2021,McClure2012,Ofek2010,Halperin2011}. On the contrary, the quantum chaos of chiral edge states are significant for their thermal equilibration in thermal transports \cite{Banerjee2018,Feldman2018,Simon2018,Ma2019}.

In this paper, we employ the exact diagonalization (ED) numerical method to explore the LSS of the many-body spectrum of a generic interacting model of two flavors of chiral complex fermions, with possible superconducting pairings, and all the irrelevant terms are ignored. The model has two analytically solvable regimes in its parameter space: the free fermion regime when the interaction is zero, and the free boson regime solvable via bosonization as a chiral Luttinger liquid. With generic parameters, we find the LSS of the interacting model in each global symmetry sector undergoes a transition from Poisson to Wigner-Dyson with respect to the global symmetry, as summarized in Fig. \ref{fig-sym}. Particularly, there is a possibly integrable regime with Poisson LSS but with no free picture, which implies the existence of hidden (quasi)local many-body conserved quantities and calls for a future analytical exploration.

The organization of the paper is as follows. We first introduce the model and its various representations in Sec. \ref{sec:model}. Next, we give its explicit eigenstate solutions in the free fermion and free boson solvable regimes in Sec. \ref{sec:solvable}. In Sec. \ref{sec-generic}, we numerically explore the LSS in generic parameter space respecting various different global symmetries, to detect the integrability and chaos of the model. We further make a comparison with the quantum chaos induced by nonlinear dispersions at high energies in Sec. \ref{sec-nonlinear}, and conclude with a discussion of open questions in Sec. \ref{sec:discussion}.


\section{The 1D chiral model}\label{sec:model}

\subsection{The complex fermion representation}

We consider a 1D model with two flavors of chiral complex fermions $c_s$ ($s=\uparrow,\downarrow$), which has an action:
\begin{equation}
S=\int dt dx \mathcal{L}(t,x)\ .
\end{equation}
The Lagrangian density takes the form
\begin{equation}
\mathcal{L}=\sum_{s=\uparrow,\downarrow}ic^\dagger_s\partial_tc_s-\mathcal{H}\ ,
\end{equation}
where the fermion fields satisfy the commutation relations 
\begin{equation}
[c_s(x),c_{s'}(x')]=\delta_{ss'}\delta(x-x'), 
\end{equation}
and the Hamiltonian density can be divided into three local terms:
\begin{equation}\label{eq:H}
\mathcal{H}=\mathcal{H}_0+\mathcal{H}_P+\mathcal{H}_I\ .
\end{equation}
The spin index $s=\uparrow,\downarrow$ here need not be the physical spin, but can be a pseudospin or any flavor index. The first term in the Hamiltonian density is a charge conserving free term of two chiral complex fermions:
\begin{equation}
\mathcal{H}_0=-i\sum_{s=\uparrow,\downarrow}v_sc^\dagger_s\partial_xc_s+\sum_{s,s'=\uparrow,\downarrow}M_{ss'}c^\dag_sc_{s'}\ ,
\end{equation}
where the velocities $v_s>0$ are real, and the matrix $M_{ss'}$ is Hermitian. 
The second term is a generic superconducting pairing term:
\begin{equation}
\mathcal{H}_P=-\frac{1}{2}\sum_{s=\uparrow,\downarrow}(iJ_s c_s\partial_x c_s+h.c.)+(\Delta c_\uparrow c_\downarrow+h.c.)\ .
\end{equation}
By a proper gauge choice, we can set the parameter $J_s$ to be real here. Lastly, there is a local (delta-function) interaction term between the local densities of the two fermion flavors:
\begin{equation}\label{eq-U}
\mathcal{H}_I=Uc^\dagger_\uparrow c_\uparrow c^\dagger_\downarrow c_\downarrow\ .
\end{equation}
The total Hamiltonian is given by $H=\int \mathcal{H}dx$. Ignoring all the irrelevant terms, this is the most generic translationally invariant interacting model for two flavors of chiral complex fermions, up to unitary transformations.

\subsection{The model rewritten with Majorana fermions}

Equivalently, one can rewrite the model of Eq. (\ref{eq:H}) in terms four flavors of chiral Majorana fermions $\psi_i$ ($i=1,2,3,4$) defined by:
\begin{equation}\label{eq-maj-basis}
c_\uparrow=\frac{\psi_1+i\psi_2}{\sqrt{2}}\ ,\qquad c_\downarrow=\frac{\psi_3+i\psi_4}{\sqrt{2}}\ .
\end{equation}
The Lagrangian density under the Majorana fermion representation can be shown to take the form
\begin{equation}
\mathcal{L}=\sum_{j=1}^4 \frac{i}{2}\psi_j\partial_t\psi_j -\mathcal{H}\ ,
\end{equation}
where the Hamiltonian density
\begin{equation}\label{eq-H-Maja}
\mathcal{H}=-\sum_{j=1}^4 \frac{i}{2}v_j\psi_j\partial_x\psi_j +\frac{i}{2}\sum_{i,j}A_{ij}\psi_i\psi_j+U\psi_1\psi_2\psi_3\psi_4\ ,
\end{equation}
where the velocities are 
\begin{equation}
\begin{split}
&v_1=v_\uparrow+J_\uparrow,\qquad v_2=v_\uparrow-J_\uparrow,\\
&v_3=v_\downarrow+J_\downarrow,\qquad v_4=v_\downarrow-J_\downarrow\ .
\end{split}
\end{equation}
and the matrix $A_{ij}$ is real antisymmetric, given by
\begin{equation}
A=\left(\begin{array}{cccc}
0&M_{\uparrow\uparrow}&\text{Im} (M_{\uparrow\downarrow}+\Delta)& \text{Re} (M_{\uparrow\downarrow}+\Delta)\\
&0&\text{Re} (\Delta-M_{\uparrow\downarrow})& \text{Im} (M_{\uparrow\downarrow}-\Delta) \\
 &&0&M_{\downarrow\downarrow}\\
 a.s.&&&0 
 \end{array}\right)\ ,
\end{equation}
where $a.s.$ stands for anti-symmetrization. 

\subsection{Bosonized representation}
The model can also be rewritten by a bosonization mapping. We define the scalar boson fields $\phi_s$ by
\begin{equation}
c_\uparrow=e^{i\phi_\uparrow}\ ,\qquad c_\downarrow=e^{i\phi_\downarrow}\ ,
\end{equation}
where the boson fields satisfy the commutation relation
\begin{equation}
[\phi_s(x),\phi_{s'}(x')]=i\pi\delta_{ss'}\text{sgn}(x-x')\ ,
\end{equation}
with $\text{sgn}(x)$ being the sign of $x$. This allows us to calculate the mapping of all operators between fermions and bosons. For instance, here we will need the mappings $c^\dag_s c_s =\frac{\partial_x\phi_s}{2\pi}$, $-ic^\dag_s\partial_x c_s =\frac{(\partial_x\phi_s)^2}{4\pi}$, and $-ic_s\partial_x c_s =2\pi e^{2i\phi_s}$ ($s=\uparrow,\downarrow$) \cite{lian2019,hu2021chiral}. As a result, our model can be mapped into a chiral boson representation
\begin{equation}
\mathcal{L}=-\frac{1}{4\pi}\sum_{s=\uparrow,\downarrow} \partial_t\phi_s\partial_x\phi_s -\mathcal{H}\ ,
\end{equation}
with the Hamiltonian density
\begin{equation}\label{eq-H-boson}
\begin{split}
\mathcal{H}&=\sum_{ss'}\frac{V_{ss'}}{4\pi}\partial_x\phi_s\partial_x\phi_s'+\sum_{s=\uparrow,\downarrow}\frac{M_{ss}}{2\pi}\partial_x\phi_s \\
&+(M_{\uparrow\downarrow}e^{i\phi_\downarrow-i\phi_\uparrow}+h.c.)+\pi \sum_{s=\uparrow,\downarrow} (J_s e^{2i\phi_s}+h.c.),
\end{split}
\end{equation}
where the velocity coefficients $V_{ss'}$ is given by
\begin{equation}\label{eq-V-matrix}
V_{\uparrow\uparrow}=v_\uparrow\ ,\ V_{\downarrow\downarrow}=v_\downarrow\ ,\ V_{\uparrow\downarrow}=V_{\downarrow\uparrow}=\frac{U}{2\pi}\ .
\end{equation}


\section{Solvable Regimes}\label{sec:solvable}

The model in Eq. (\ref{eq:H}) has two solvable cases, which give free chiral fermions and free chiral bosons (Luttinger liquid), respectively. We discuss these two solvable cases in this section.

\subsection{The case of free chiral Majorana fermions}\label{sec:freefermion}
When the fermion interaction vanishes, namely, 
\begin{equation}
U=0\ ,
\end{equation}
one simply has free fermions. In the Majorana fermion representation, we define the momentum space Majorana fermions
\begin{equation}\label{eq-Maj-modes}
\psi_{j,k}=\frac{1}{\sqrt{L}}\int e^{-ikx}\psi_j(x)\ ,\qquad \psi_{j,-k}=\psi_{j,k}^\dag\ ,
\end{equation}
where $L$ is the spatial length of the system. Without bulk flux insertion, the fermions satisfy anti-periodic boundary conditions, thus the single-fermion momentum $k\in\frac{2\pi}{L}(\mathbb{Z}+\frac{1}{2})$. By defining $\psi_k=(\psi_{1,k},\psi_{2,k},\psi_{3,k},\psi_{4,k})^T$, one can then rewrite the Hamiltonian in Eq. (\ref{eq-H-Maja}) with $U=0$ as
\begin{equation}
H=\int \mathcal{H}dx =\frac{1}{2}\sum_{k}\psi_{-k}^T h(k)\psi_{k}\ ,
\end{equation}
where the $4\times4$ matrix $h(k)$ is given by
\begin{equation}\label{eq-free-h}
h_{ij}(k)=\delta_{ij}v_j k +i A_{ij}\ .
\end{equation}
Diagonalizing the matrix $h(k)$ then gives the single-fermion energy spectrum $\epsilon_n(k)$ ($1\le n\le 4$) of the model. We emphasize that for the system to have a lower energy bound and be stable, the parameters have to satisfy $v_j\ge0$ ($1\le j\le 4$).

\subsection{The case of free chiral bosons with U(1)$^{(\uparrow)}\times$U(1)$^{(\downarrow)}$ symmetry}\label{sec:freeboson}
The other solvable point is the chiral Luttinger liquid point, which is when each of the fermion spin flavor has a U(1) charge symmetry, namely, when the system has a total global symmetry U(1)$_\uparrow\times$U(1)$_\downarrow$. This requires the vanishing of the following parameters:
\begin{equation}\label{eq-boson-condition}
J_s=0\ ,\quad \Delta=0\ , \quad M_{\uparrow\downarrow}=0\ .
\end{equation}
Therefore, there is no superconductivity pairing, i.e., $\mathcal{H}_P=0$. By Eq. (\ref{eq-H-boson}), the bosonized Hamiltonian becomes a free boson Hamiltonian with terms no higher than the second order of boson fields $\phi_s$, although the fermion form of the Hamiltonian is interacting. The boson fields $\phi_s$ ($s=\uparrow,\downarrow$) can be expanded in modes as
\begin{equation}
\phi_s(x)=\phi_{0,s}+\frac{2\pi N_s}{L} x +\sum_{k>0}(a_{s,k}e^{ikx}+a_{s,k}^\dag e^{-ikx})\ ,
\end{equation}
where $L$ is the spatial length, and $a_{s,k}$ and $a_{s,k}^\dag$ are the annihilation and creation operators of the normal boson modes. Besides,
\begin{equation}
N_s=\int \normord{c_s^\dag(x) c_s(x)} dx
\end{equation}
is the U(1) charge (or number of fermions) of the spin $s$ (where $\normord{O}$ stands for normal ordering of operator $O$), and it satisfies the commutation relation $[\frac{2\pi N_s}{L},\phi_{0,s}]=i$ with the constant piece $\phi_{0,s}$. If we impose the anti-periodic boundary condition for the fermions, the bosons will satisfy periodic boundary condition, and their momenta take values $k\in\frac{2\pi}{L}\mathbb{Z}$. This leads to a free boson Hamiltonian
\begin{equation}
H=\frac{\pi}{L}\sum_{ss'} V_{ss'}N_sN_{s'} +\sum_{s=\uparrow,\downarrow}M_{ss} N_s +\sum_{\eta=\pm}\sum_{k>0}v_{\eta}k b_{\eta,k}^\dag b_{\eta,k}\ ,
\end{equation}
where we have defined $v_\eta$ ($\eta=\pm$) as the eigenvalues of the matrix $V_{ss'}$ in Eq. (\ref{eq-V-matrix}), and new boson eigenmodes $b_{\eta,k}$:
\begin{equation}\label{eq-eigenboson}
\sum_{s'}V_{ss'} \zeta_{s'\eta}=v_\eta\zeta_{s\eta}\ ,\quad b_{\eta,k}=\sum_s\zeta_{s\eta}^*a_{s,k}\ .
\end{equation}
This is known as the chiral Luttinger liquid, where the model reduces to two free chiral boson modes with velocities $v_\pm$. We note that the stability of the system requires $v_\pm\ge0$, which avoids infinite negative energy states.

We note that if $v_\uparrow=v_\downarrow$, one can relax the condition in Eq. (\ref{eq-boson-condition}) to allow nonzero $M_{\uparrow\downarrow}$, and still gets free chiral bosons. This is because in this case, both the fermionic velocity kinetic term $-i\sum_{s}v_sc^\dagger_s\partial_xc_s$ and the interaction term $Uc^\dagger_\uparrow c_\uparrow c^\dagger_\downarrow c_\downarrow$ are invariant under any SU(2) fermion basis rotation. One can therefore rotate the fermion basis $(c_\uparrow,c_\downarrow)^T$ to a new basis $(c_\uparrow',c_\downarrow')^T=\mathcal{U}(c_\uparrow,c_\downarrow)^T$ which diagonalizes the $M$ matrix. In this new basis, one again satisfies condition (\ref{eq-boson-condition}), and can thus bosonize the model into free chiral bosons.

\section{Generic Parameters: An Exact Diagonalization Study}\label{sec-generic}

With generic parameters, the model is no longer free in either the fermion or the boson representations, thus there is no obvious analytical ways to solve it. Therefore, we numerically calculate its eigenstates and energy spectrum by exact diagonalization (ED). For this purpose, we numerically construct and diagonalize the Hamiltonian in its fermion representation. We impose anti-periodic boundary condition in the spatial direction, and set the spatial length to $L=2\pi$ without loss of generality. Accordingly, all the single-fermion momenta are half-odd integers, namely, 
\begin{equation}\label{eq-k-half}
k\in\mathbb{Z}+\frac{1}{2}\ .
\end{equation}

The many-body total momentum $K_{\text{tot}}$ is always conserved and nonnegative. From the chiral Majorana fermion representation in Eq. (\ref{eq-Maj-modes}), it is clear that all the Majorana fermion modes have positive momentum. Thus, for a fixed total momentum $K_{\text{tot}}$, the many-body Hilbert space dimension is finite, since the allowed number of fermions are upper bounded. This makes the ED study of the model possible.

In the below, we investigate the numerical spectrum of the interacting model in Eq. (\ref{eq:H}) under different symmetry constraints of the parameters, to examine whether the model is integrable or chaotic. Generically, we assume the two spin flavors have different free velocities (unless specified), namely, $v_\uparrow\neq v_\downarrow$. We will show that as the global symmetry lowers, the model exhibits a transition from quantum integrable to many-body quantum chaotic.

\subsection{Probing quantum chaos}

A well-known diagnostics of quantum chaos is the many-body level spacing statistics (LSS) in a conserved symmetry charge sector. In this paper, by the symmetry charges we refer to those of the global symmetries of the model. 

In particular, the generic model in Eq. (\ref{eq:H}) always has two conserved symmetry charges: the total fermion parity $(-1)^{N_\uparrow+N_\downarrow}$, and the total many-body momentum $K_\text{tot}$ from the translation symmetry. Since we imposed anti-periodic boundary condition, all the single-fermion momenta are half-odd integers (Eq. (\ref{eq-k-half})), and thus the two conserved charges are not independent:
\begin{equation}\label{eq-total-parity}
(-1)^{N_\uparrow+N_\downarrow}=(-1)^{2K_{\text{tot}}}\ .
\end{equation}
Therefore, it is sufficient to keep only the total momentum $K_{\text{tot}}$ for the above two conserved charges.

\begin{figure}[tbp]
\begin{center}
\includegraphics[width=3.3in]{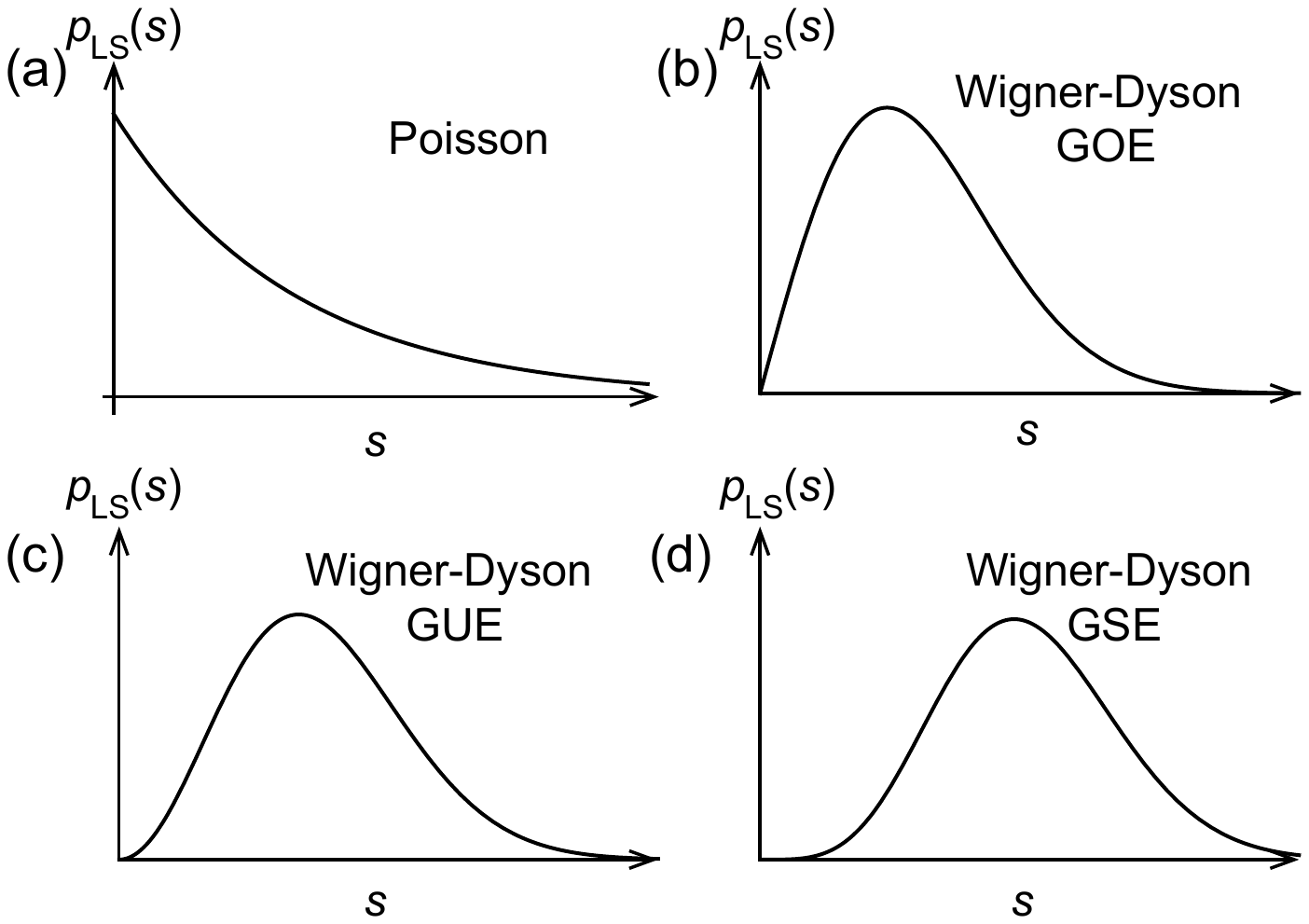}
\end{center}
\caption{Illustration of different kinds of level spacing statistics.
}
\label{fig-gLSS}
\end{figure}

Assume the $n$-th many-body energy level (sorted from the lowest to the highest) in a conserved charge sector $Q$ (which includes momentum $K_\text{tot}$) is $E_{n}(Q)$. One can define the level spacing $\delta_{E,n}=E_{n+1}(Q)-E_{n}(Q)$, and examine the statistical probability distribution $p_{LS}(\delta_E)$ of $\delta_{E,n}$, which is known as the LSS. There are generically two situations:

(i) If the system is quantum integrable (exactly solvable), or if there are still hidden (quasi)local conserved quantities in the conserved charge sectors $Q$, the LSS in sector $Q$ will resemble the \emph{Poisson distribution} (characterizing independent random variables) \cite{Berry1977}:
\begin{equation}\label{eq-Poisson}
p_{LS}(s)\propto e^{-s/s_0}\ ,
\end{equation}
where the constant $s_0\ge0$ (see Fig. \ref{fig-gLSS}(a)). This indicates there is no repulsion between neighboring energy levels.

(ii) If the system is fully quantum chaotic within a conserved charge sector $Q$, the LSS in sector $Q$ will resemble the LSS of random Hermitian matrices $H$, which is known as the \emph{Wigner-Dyson distribution} \cite{Bohigas1984,Dyson1970,Wigner1967}. Depending on symmetry classes of the system, the Wigner-Dyson distribution function is given by
\begin{equation}\label{eq-WD}
p_{LS}(s)\propto s^me^{-s^2/s_0^2}\ ,
\end{equation}
where $s_0>0$. The integer $m=1,2,4$ for the Hamiltonian $H$ in the real (spinless time-reversal (TR) invariant), complex (without TR invariance) and symplectic (spinful TR invariant with spin-orbit coupling), respectively. Note that $p_{LS}(0)=0$ in this case, indicating that the neighboring energy levels repulse each other.

We will also numerically compute the zero-temperature spectral weight $A_s(\omega,k)=2\text{Im}G_{R,s}(\omega,k)$ of fermions $c_s$, where $G_{R,s}(\omega,k)$ is the retarded Green's function of fermion $c_s$ in the energy-momentum space. If $|k,j\rangle$ denotes the $j$-th many-body eigenstate with total momentum $K_\text{tot}=k$ and energy $E_{k,j}$, where $1\le j\le N_k$ with Hilbert space dimension $N_k$ ($N_k>0$ if and only if $k>0$), and $|0\rangle$ denotes the zero-particle vacuum state, we can numerically calculate the spectral weight as
\begin{equation}\label{eq-SW}
\begin{split}
A_{s}(\omega,k)=&\sum_{j=1}^{N_{-k}}|\langle-k,j|c_s(k)|0\rangle|^2\delta(\omega+E_{-k,j})\\
&+\sum_{j=1}^{N_{k}}|\langle 0|c_s(k)|k,j\rangle |^2\delta(\omega-E_{k,j})\ .
\end{split}
\end{equation}
As is clear from this expression, the spectral weight characterizes the single-fermion density of states. In practical calculations, to avoid numerical divergences, we relax the delta function into a Lorentzian function 
\begin{equation}
\delta(\omega)\rightarrow  \frac{1}{\pi}\frac{\eta}{\omega^2+\eta^2}\ ,
\end{equation}
and take $\eta=0.3$.

\subsection{The free fermion and free boson solvable points}

We first examine the numerical LSS at the 2 solvable points of free fermions and free bosons we discussed in Sec. \ref{sec:solvable}. As free models, they are many-body quantum integrable, since all the many-body states are Fock states of the single-particle eigenstates. Therefore, one expects the LSS of their many-body energy spectrum in each charge sector to show Poisson distributions.

\begin{figure}[tbp]
\begin{center}
\includegraphics[width=3.3in]{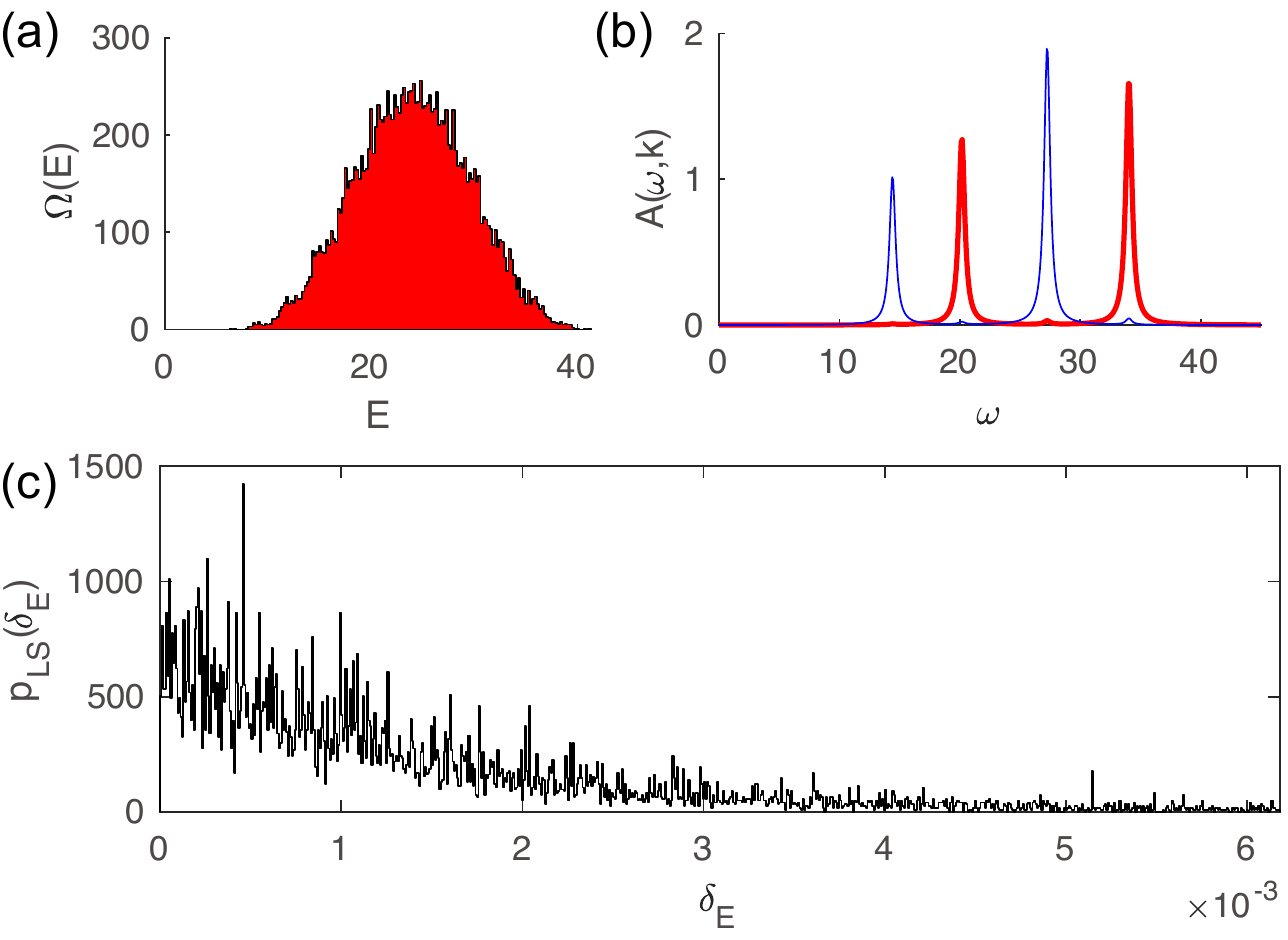}
\end{center}
\caption{The ED results for the free fermion case, where the parameters are given by $U=0$,  $v_\uparrow=2$, $v_\downarrow=1.55$, $M_{\uparrow\uparrow}=1$, $M_{\downarrow\downarrow}=1$, $M_{\uparrow\downarrow}=1+0.5i$, $J_\uparrow=0.5$, $J_\downarrow=0.45$, and $\Delta=0.5+0.3i$. The panels show (a) the DOS of the total momentum $K_\text{tot}=\frac{27}{2}$ sector; (b) The spectral weights $A_s(\omega,k)$ of the spin up (red thick line) and spin down (blue thin line) fermions at $k=\frac{27}{2}$; and (c) the LSS of the total momentum $K_\text{tot}=\frac{27}{2}$ sector.
}
\label{fig-free}
\end{figure}

\emph{The free fermion case}. In this case with the interaction $U=0$ in Eq. (\ref{eq-U}), and all the other parameters nonzero (Sec. \ref{sec:freefermion}), there are only the total fermion parity $\mathbb{Z}_2$ symmetry and the translational symmetry, the conserved charges of which are the parity $(-1)^{N_\uparrow+N_\downarrow}$ and the total many-body momentum $K_{\text{tot}}$. As shown in Eq. (\ref{eq-total-parity}), these two conserved charges are not independent, and we can label each symmetry charge sector by $K_\text{tot}$.

Fig. \ref{fig-free}(a) and (c) shows the many-body density of states (DOS) and LSS of the free fermion case ($U=0$) in the sector of total momentum $K_\text{tot}=\frac{27}{2}$. The other parameters are listed in the caption of Fig. \ref{fig-free}, which are chosen sufficiently arbitrary so that there are no additional global symmetries. As one can easily see, the LSS shows a Poisson statistics, due to the many-body integrable nature of free fermions. 

Fig. \ref{fig-free}(b) shows the zero-temperature spectral weights of $c_\uparrow$ (red thick line) and $c_\downarrow$ (blue thin line) at momentum $k=K_\text{tot}=\frac{27}{2}$, which are defined in Eq. (\ref{eq-SW}). As expected, they show delta function peaks at the single-particle energies of the free chiral Majorana fermions (eigenvalues of Eq. (\ref{eq-free-h})).

\emph{The free boson case}. As shown in Sec. \ref{sec:freeboson}, when $J_s=0$, $\Delta=0$, and $M_{\uparrow\downarrow}=0$, while the interaction $U\neq0$, the model has a global symmetry U(1)$_\uparrow\times$U(1)$_\downarrow$, and is solvable as two flavors of free chiral bosons. The conserved symmetry charges are therefore $N_\uparrow$ and $N_\downarrow$.

\begin{figure}[tbp]
\begin{center}
\includegraphics[width=3.3in]{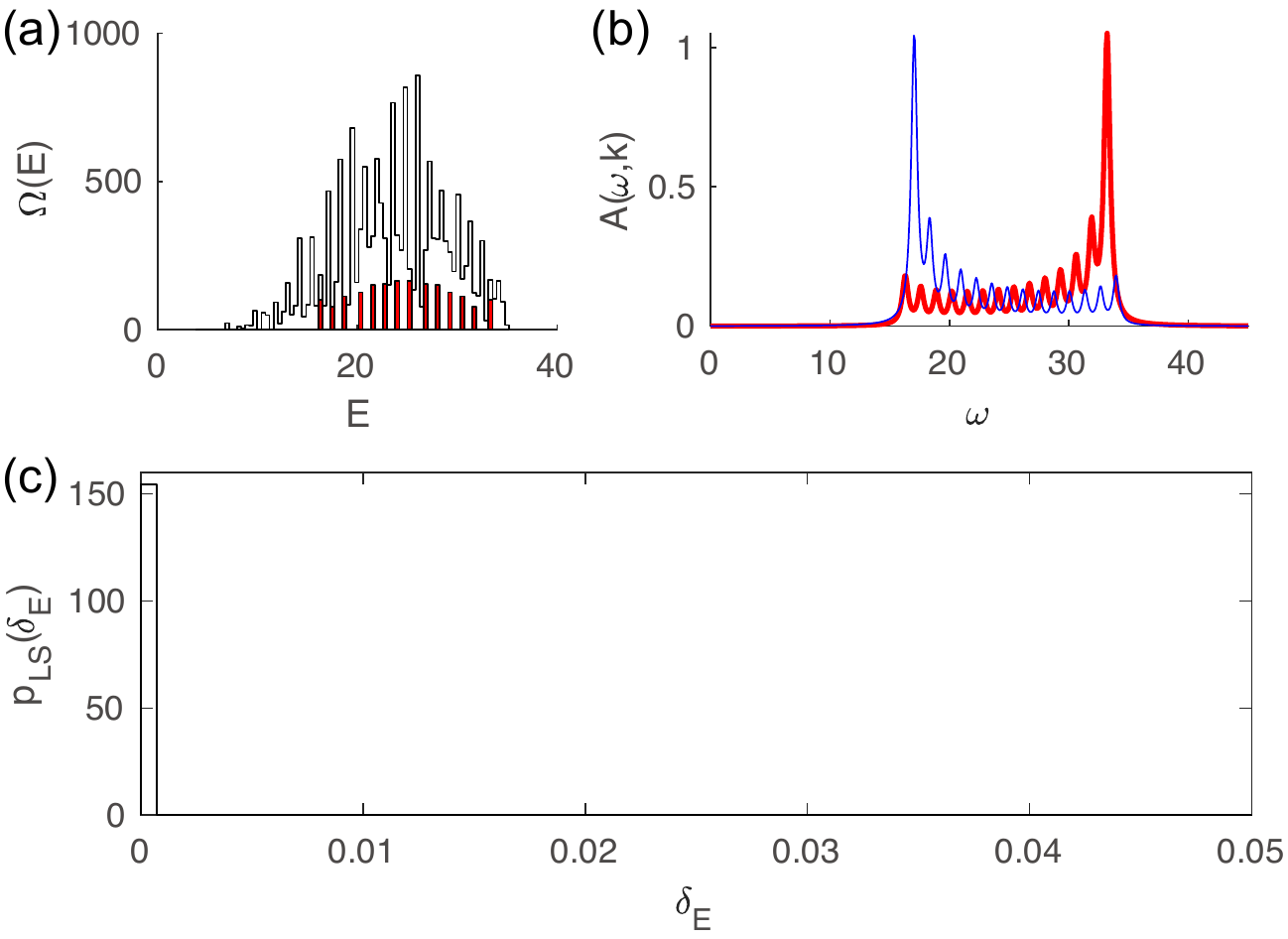}
\end{center}
\caption{ED calculation for the free boson case (with U(1)$_\uparrow\times$U(1)$_\downarrow$ symmetry), where the parameters are $U=1.2\pi$,  $v_\uparrow=2$, $v_\downarrow=1.5$, $M_{\uparrow\uparrow}=1$, $M_{\downarrow\downarrow}=2$, $M_{\uparrow\downarrow}=0$, $J_\uparrow=0$, $J_\downarrow=0$, and $\Delta=0$. The panels show (a) the DOS of the total momentum $K_\text{tot}=\frac{27}{2}$ sector (unfilled line) and its subsector with $N_\uparrow=1$, $N_\downarrow=0$ (the line filled with red); (b) The spectral weights $A_s(\omega,k)$ of the spin up (red thick line) and spin down (blue thin line) fermions at $k=\frac{27}{2}$; and (c) the LSS of the symmetry sector of $(K_\text{tot}=\frac{27}{2},N_\uparrow=1,N_\downarrow=0)$ (red part in (a)).
}
\label{fig-U1U1}
\end{figure}

Fig. \ref{fig-U1U1} shows the ED results of such a free-boson example (parameters given in the caption). In Fig. \ref{fig-U1U1}(a), the unfilled line is the total DOS of the total momentum $K_\text{tot}=\frac{27}{2}$ sector; while the line filled with red color is the DOS of the finer symmetry sector with quantum numbers $(K_\text{tot}=\frac{27}{2},N_\uparrow=1,N_\downarrow=0)$. Fig. \ref{fig-U1U1}(c) shows the LSS of this symmetry sector $(K_\text{tot}=\frac{27}{2},N_\uparrow=1,N_\downarrow=0)$, which is almost a delta function at zero. This is because the free bosons' linear dispersion leads to a large number of many-body level degeneracy. Nonetheless, the LSS can be viewed as a Poisson distribution with a large slope (i.e., small $s_0$ in Eq. (\ref{eq-Poisson})).

Fig. \ref{fig-U1U1}(b) shows the spectral weights of the spin up (red thick line) and down (blue thin line) fermions in this free boson case, respectively. Analytically, by refermionization, one can derive the spectal weights of the spin $s$ fermion as \cite{lian2019,hu2021chiral,hu2021integrability} (up to energy shifts induced by the chemical potentials $M_{\uparrow\uparrow}$ and $M_{\downarrow\downarrow}$)
\begin{equation}
A_{s}(\omega,k)=\frac{2\Theta(\omega-v_-k)\Theta(v_+k-\omega)}{(\omega-v_-k)^{1-|\zeta_{s+}|^2}(v_+k-\omega)^{|\zeta_{s+}|^2}}\ ,
\end{equation}
where $\zeta_{s\eta}$ are the coefficients in Eq. (\ref{eq-eigenboson}). This agrees well with the numerical results in Fig. \ref{fig-U1U1}(b).

\subsection{The case with U(1) symmetry}\label{subsec-U1}

We now consider the case of adding a nonzero hopping $M_{\uparrow\downarrow}$ to the solvable free-boson point, namely,
\begin{equation}\label{eq-U1-condition}
J_s=0\ ,\quad \Delta=0\ , \quad M_{ss'}\neq 0, \quad U\neq0\ .
\end{equation}
Due to the nonzero term $M_{\uparrow\downarrow}$, the model only has a global U(1) symmetry. Thus, the conserved symmetry charges are the total fermion charge $N=N_\uparrow+N_\downarrow$ and the total momentum $K_\text{tot}$.

In addition, in the special case when $M_{\uparrow\uparrow}=M_{\downarrow\downarrow}$, the model Hamiltonian obeys a simple transformation under the particle-hole transformation $P$ that flips the U(1) fermion charge $N$:
\begin{equation}
\begin{split}
&P c_s P^{-1}=e^{i\theta_s}c_s^\dag,\quad P c_s^\dag P^{-1}=e^{-i\theta_s}c_s, \\
&P N_s P^{-1}=-N_s, \\
&P H P^{-1}=H-2M_{\uparrow\uparrow}N\ ,
\end{split}
\end{equation}
where $\theta_s=s(\arg(M_{\uparrow\downarrow})+\frac{\pi}{2})$ for $s=\pm$ (corresponding to $s=\uparrow,\downarrow$). In this case (namely, $M_{\uparrow\uparrow}=M_{\downarrow\downarrow}$), the $N=0$ charge sector will have an additional symmetry $P$ and thus an additional conserved charge, the eigenvalue $\eta_P=\pm1$ of operator $P$. The $N\neq0$ sectors do not have this additional conserved charge.

Although the model takes a simple form, it cannot be solved as free bosons or free fermions. In the fermion representation, the interaction $U$ makes it not free. In the bosonized representation, the $M_{\uparrow\downarrow}$ term is bosonized into a nonlinear term
\begin{equation}
M_{\uparrow\downarrow} e^{i\phi_\uparrow-i\phi_\downarrow}+h.c.\ ,
\end{equation}
making the bosons not free, either. It is not yet known if the model is exactly solvable by certain many-body techniques. Therefore, instead, we examine the ED results of the model in this case.

\begin{figure}[tbp]
\begin{center}
\includegraphics[width=3.3in]{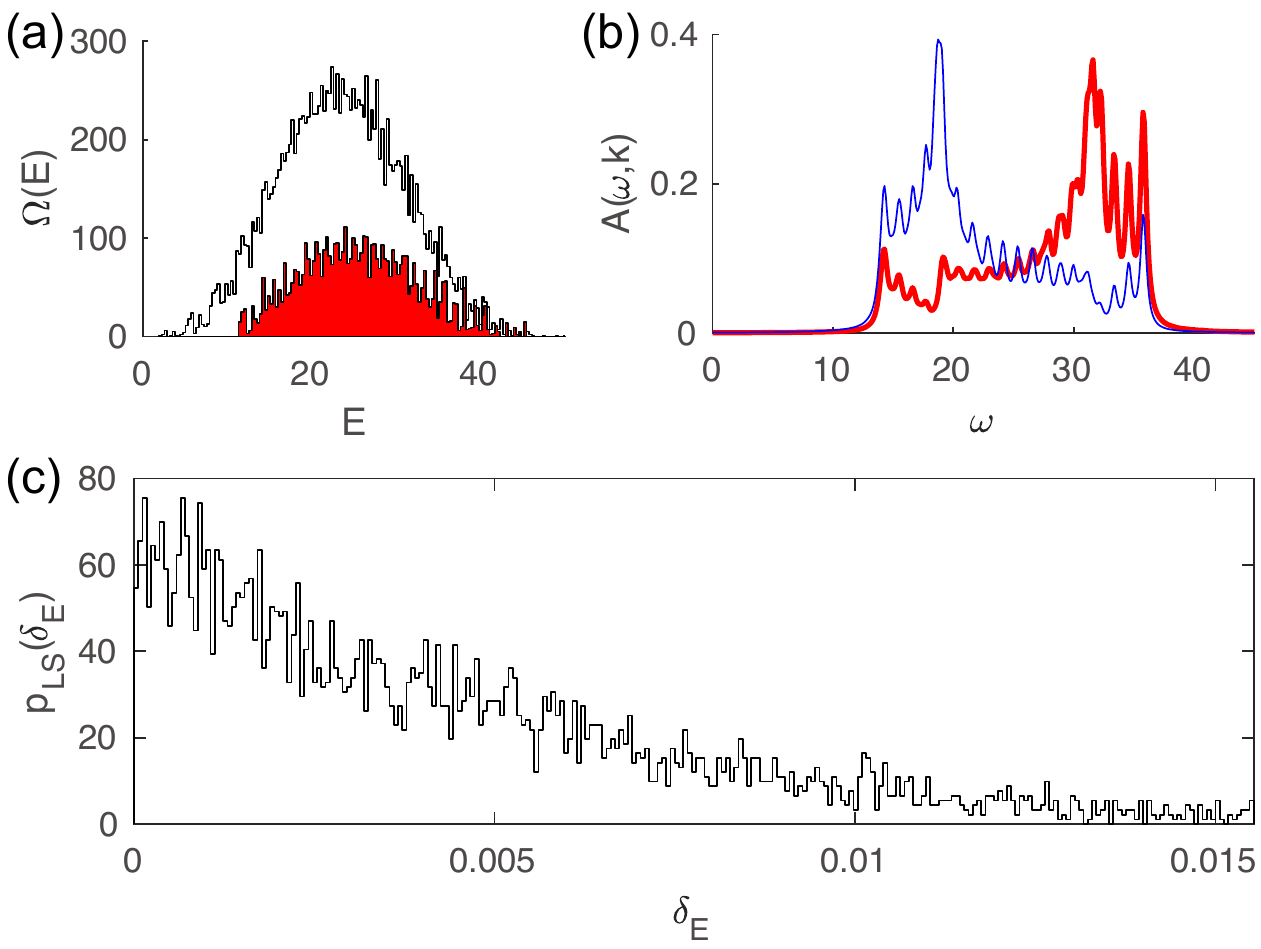}
\end{center}
\caption{ED calculation for the case with U(1) symmetry, with parameters $U=1.2\pi$,  $v_\uparrow=2$, $v_\downarrow=1.5$, $M_{\uparrow\uparrow}=1$, $M_{\downarrow\downarrow}=2$, $M_{\uparrow\downarrow}=2.4$, $J_\uparrow=0$, $J_\downarrow=0$, and $\Delta=0$. Note that $M_{\uparrow\downarrow}$ can always be taken as real via a relative U(1) rotation between the two spins. (a) the DOS of the $K_\text{tot}=\frac{27}{2}$ sector (unfilled line) and the subsector with total fermion charge $N=1$ (the line filled with red); (b) The spectral weights $A_s(\omega,k)$ of the spin up (red thick line) and spin down (blue thin line) fermions at $k=\frac{27}{2}$; and (c) the LSS of the symmetry sector of $(K_\text{tot}=\frac{27}{2},N=1)$ (red part in (a)).
}
\label{fig-U1}
\end{figure}

Fig. \ref{fig-U1} shows the numerical results for a set of arbitrarily chosen parameters (given in Fig. \ref{fig-U1} caption) in this case. Panel (a) shows the DOS of the $K_\text{tot}=\frac{27}{2}$ sector (the unfilled line), and its subsector with symmetry charges with $(K_\text{tot}=\frac{27}{2},N=1)$ (line filled with red color), which is much smoother compared to the free boson case (Fig. \ref{fig-U1U1}(a)). The fermion spectral weights in Fig. \ref{fig-U1}(b) show irregular shapes, which is an indication of the absence of free fermion or free boson picture. 

Intriguingly, the LSS in the finest symmetry sector in this case still shows Poisson statistics among the parameter space we have explored. Fig. \ref{fig-U1}(c) shows the LSS in the symmetry sector of $(K_\text{tot}=\frac{27}{2},N=1)$ (i.e., the red part of DOS in Fig. \ref{fig-U1}(a)). We have examined the LSS in different symmetry sectors for more sets of parameters satisfying Eq. (\ref{eq-U1-condition}), all of which show no level repulsions. This indicates the existence of hidden (quasi)local conserved quantities \cite{lian2022conserv} beyond those of the global symmetries, which has not been theoretically understood yet. Moreover, the model in this case may be even totally quantum integrable, which we leave for the future studies. 

Another similar case with different symmetries will be presented in Sec. \ref{subsec-U1Z2} below. Lastly, we note that when we set $U=0$ in this case, the model will become free fermions with nonlinear dispersions (due to the nonzero $M_{ss'}$). However, the behavior of the LSS with $U\neq0$ here (Poisson) is completely different from that of generic nonlinear dispersion fermions with interaction (which is Wigner-Dyson, see Sec. \ref{sec-nonlinear}).

\subsection{The case with U(1)$^{(\uparrow)}\times\mathbb{Z}_{2}^{(\downarrow)}$ symmetry}\label{subsec-U1Z2}

In this subsection, we investigate the case with a superconducting pairing: starting from the free boson case with global symmetry U(1)$_\uparrow\times$U(1)$_\downarrow$, we turn on the pairing within the spin down flavor, reducing the global symmetry into U(1)$^{(\uparrow)}\times\mathbb{Z}_{2}^{(\downarrow)}$. The parameters thus satisfy
\begin{equation}\label{eq-U1Z2-condition}
J_\uparrow=0\ ,\quad J_\downarrow\neq0\ ,\quad  \Delta=0\ , \quad M_{\uparrow\downarrow}=0\ , \quad U\neq0\ .
\end{equation}
Accordingly, the conserved charges are $K_\text{tot}$ and $N_\uparrow$. The parity $(-1)^{N_\downarrow}$ is dependent on $K_\text{tot}$ and $N_\uparrow$, as we showed in Eq. (\ref{eq-total-parity}).

A special case within the parameter space of Eq. (\ref{eq-U1Z2-condition}) is when $M_{\downarrow\downarrow}=0$, for which the model transforms simply under a particle-hole transformation $P$:
\begin{equation}
\begin{split}
&P c_s P^{-1}=c_s^\dag,\quad P c_s^\dag P^{-1}=c_s,\quad P N_s P^{-1}=-N_s, \\
&P H P^{-1}=H-2M_{\uparrow\uparrow}N_\uparrow\ .
\end{split}
\end{equation}
Accordingly, when $M_{\downarrow\downarrow}=0$, the $N_\uparrow=0$ charge sector has an additional symmetry $P$, and thus gains an additional conserved charge, the eigenvalue $\eta_P=\pm1$ of the operator $P$. This additional charge does not exist in all the $N_\uparrow\neq0$ sectors.

\begin{figure}[tbp]
\begin{center}
\includegraphics[width=3.3in]{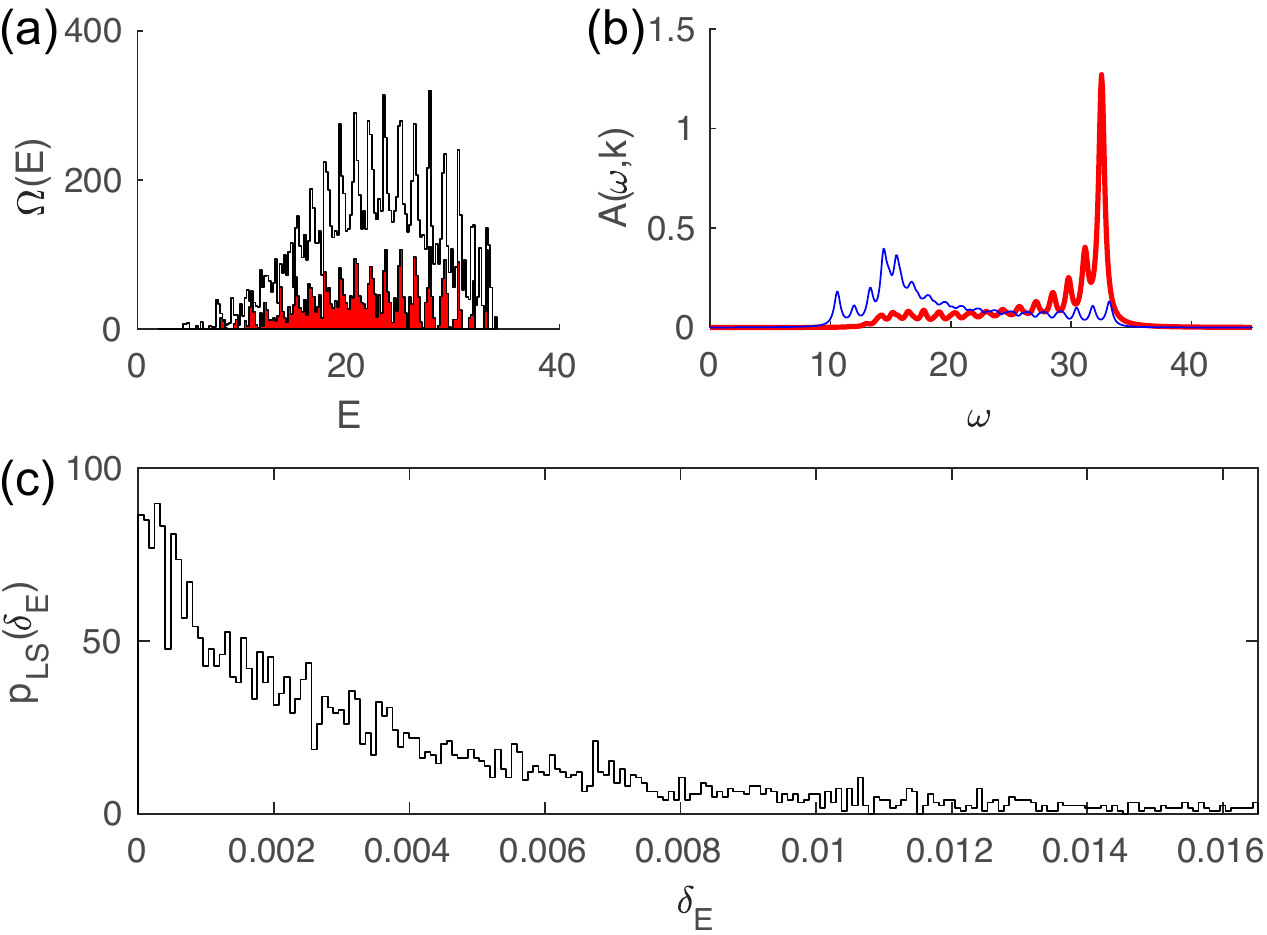}
\end{center}
\caption{ED calculation in the case with U(1)$^{(\uparrow)}\times\mathbb{Z}_{2}^{(\downarrow)}$ global symmetry. The parameters are $U=1.2\pi$,  $v_\uparrow=2$, $v_\downarrow=1.3$, $M_{\uparrow\uparrow}=1$, $M_{\downarrow\downarrow}=2$, $M_{\uparrow\downarrow}=0$, $J_\uparrow=0$, $J_\downarrow=0.2$, and $\Delta=0$. (a) the DOS of the $K_\text{tot}=\frac{27}{2}$ sector (unfilled line) and the subsector with total fermion charge $N_\uparrow=0$ and thus $(-1)^{N_\downarrow}=(-1)^{N_\uparrow+2K_\text{tot}}=-1$ (the line filled with red). (b) The spectral weights $A_s(\omega,k)$ of spin up (red thick line) and spin down (blue thin line) fermions at $k=\frac{27}{2}$. (c) The LSS of the symmetry sector of $(K_\text{tot}=\frac{27}{2},N_\uparrow=0)$ (red part in (a)).
}
\label{fig-U1Z2}
\end{figure}

Such a U(1)$^{(\uparrow)}\times\mathbb{Z}_{2}^{(\downarrow)}$ symmetry constraint may seem unphysical, that different spins have different symmetries. However, if one regard the spin solely as a fermion flavor index, the model can have its physical context. For instance, it is shown in Ref. \cite{hu2021integrability} that the interacting chiral edge states of the $4/3$ filling FQH state are equivalent to the interacting fermion model with U(1)$^{(\uparrow)}\times\mathbb{Z}_{2}^{(\downarrow)}$ here, where the spin $\uparrow$ fermion carries an irrational electric charge $\frac{2e}{\sqrt{3}}$, while the spin $\downarrow$ fermion is charge neutral. Thus, with charge conservation, pairing is allowed for spin $\downarrow$ fermions, but not allowed for spin $\uparrow$ fermions.

With the pairing term $J_\downarrow$, the model has a nonlinear term 
\begin{equation}
J_\downarrow e^{2i\phi_\downarrow}+h.c. 
\end{equation}
in the bosonized representation. Therefore, the model is neither a free fermion nor a free boson model. Intriguingly, as studied in Ref. \cite{hu2021integrability}, this model with U(1)$^{(\uparrow)}\times\mathbb{Z}_{2}^{(\downarrow)}$ symmetry, i.e., with parameters satisfying Eq. (\ref{eq-U1Z2-condition}), shows Poisson LSS in each global symmetry charge sector. Fig. \ref{fig-U1Z2} shows the ED results of an example, with the parameters as given in the Fig. \ref{fig-U1Z2} caption. The unfilled line and red-filled line in Fig. \ref{fig-U1Z2}(a) are the DOS of the entire $K_\text{tot}=\frac{27}{2}$ sector and the DOS of the subsector with $(K_\text{tot}=\frac{27}{2},N_\uparrow=0)$, respectively. As expected, the fermion spectral weights in Fig. \ref{fig-U1Z2}(b) are different from those in the free fermion or free boson cases. Fig. \ref{fig-U1Z2}(c) shows the LSS in the symmetry sector $(K_\text{tot}=\frac{27}{2},N_\uparrow=0)$, which is a clear Poisson distribution. More symmetry sectors are examined in Ref. \cite{hu2021integrability}, all of which shows a Poisson LSS.

Therefore, similar to the U(1) symmetry case we discussed in Sec. \ref{subsec-U1}, the Poisson distribution indicates the existence of hidden (quasi)local conserved quantities \cite{lian2022conserv}, and moreover, the model may be fully quantum integrable. Identifying such hidden conserved quantities is an interesting task for the future studies.

We comment on an observation, that in both the U(1) symmetric case in Sec. \ref{subsec-U1} and the U(1)$^{(\uparrow)}\times\mathbb{Z}_{2}^{(\downarrow)}$ case here, the bosonized representation of the model has only one nonlinear sine or cosine term in the boson fields $\phi_s$: the $M_{\uparrow\downarrow}$ term in the former case, and the $J_\downarrow$ term in the later case. This may be intrinsically related to their Poisson LSS and potential quantum integrability. As we will see in the next few subsections, if one has two or more nonlinear sine or cosine terms, the LSS in each symmetry sector will show Wigner-Dyson statistics.

\subsection{The case with $\mathbb{Z}_{2}^{(\uparrow)}\times\mathbb{Z}_{2}^{(\downarrow)}\times\mathbb{Z}_{2}^{(++)}$ symmetry}\label{subsec-Z2Z2Z2}

\begin{figure}[tbp]
\begin{center}
\includegraphics[width=3.3in]{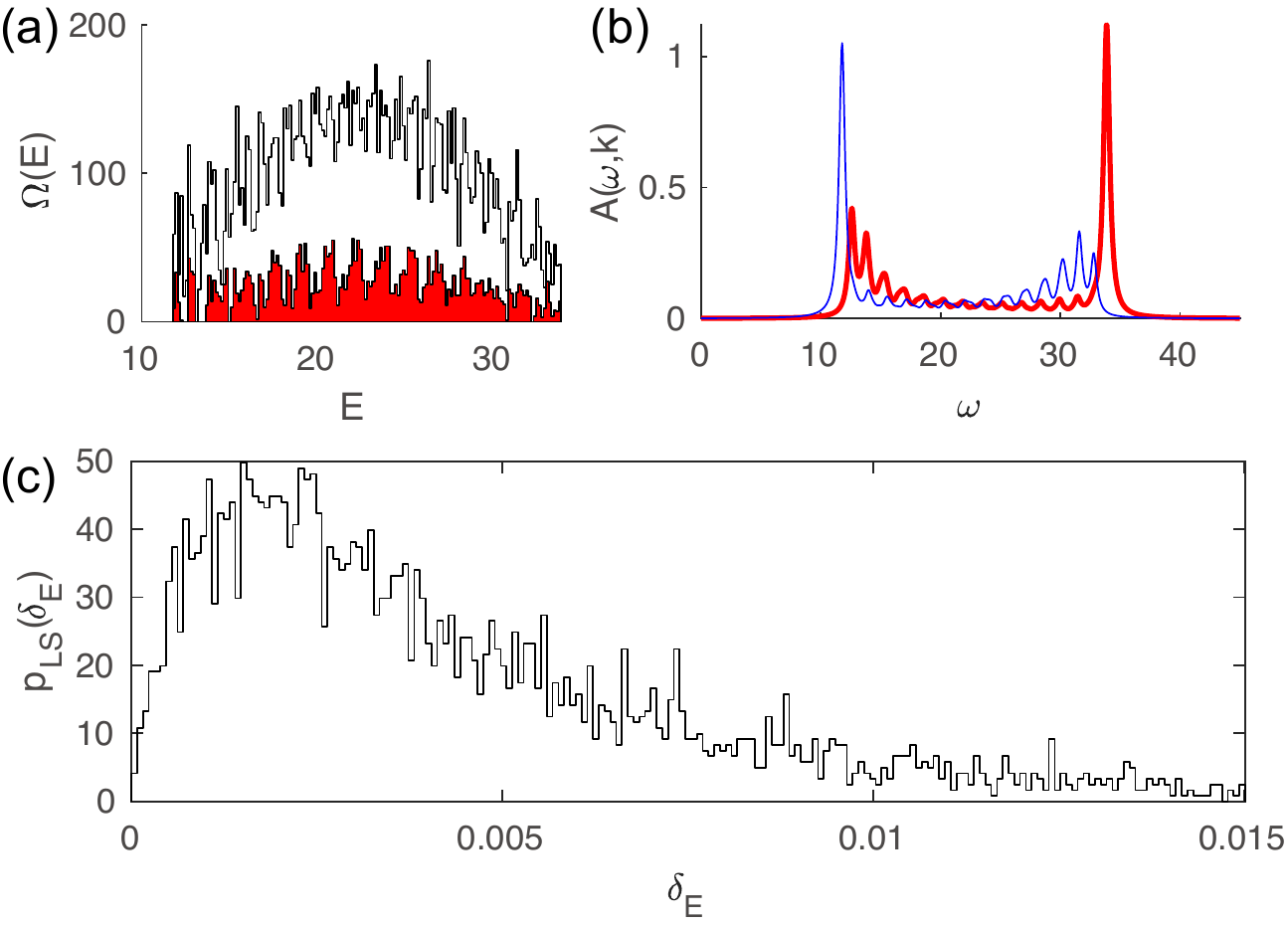}
\end{center}
\caption{ED calculation for the case with $\mathbb{Z}_{2}^{(\uparrow)}\times\mathbb{Z}_{2}^{(\downarrow)}\times\mathbb{Z}_{2}^{(++)}$ symmetry, where the parameters are $U=1.2\pi$,  $v_\uparrow=1.8$, $v_\downarrow=1.55$, $M_{\uparrow\uparrow}=0$, $M_{\downarrow\downarrow}=0$, $M_{\uparrow\downarrow}=0$, $J_\uparrow=0.5$, $J_\downarrow=0.45$, and $\Delta=0$. (a) The DOS of the $K_\text{tot}=\frac{27}{2}$ sector (unfilled line) and the subsector with $(-1)^{N_\uparrow}=+1$, $P_{++}=+1$, and thus $(-1)^{N_\downarrow}=(-1)^{N_\uparrow+2K_\text{tot}}=-1$ (the line filled with red). (b) The spectral weights $A_s(\omega,k)$ of spin up (red thick line) and spin down (blue thin line) fermions at $k=\frac{27}{2}$. (c) The LSS of the symmetry sector of $(K_\text{tot}=\frac{27}{2},(-1)^{N_\uparrow}=+1,P_{++}=+1)$ (red part in (a)).
}
\label{fig-Z2Z2Z2}
\end{figure}

We now turn on more pairing terms in our model, and examine its many-body LSS. In this subsection, we consider parameters satisfying
\begin{equation}\label{eq-Z2Z2Z2-condition}
J_s\neq0\ ,\ \ \Delta=0\ , \ \ M_{\uparrow\uparrow}=M_{\downarrow\downarrow}=M_{\uparrow\downarrow}=0\ ,\ \  U\neq0\ ,
\end{equation}
where both spin up and spin down have a nonzero p-wave pairing $J_s$. It is straightforward to see that each spin $s$ has a fermion parity symmetry $\mathbb{Z}_2^{(s)}$, with conserved charges $(-1)^{N_s}$, respectively. Moreover, the fact that all the mass terms $\Delta$, $M_{ss'}$ are zero leads to another implicit parity symmetry, which we call $\mathbb{Z}_2^{(++)}$. The parity charge of this $\mathbb{Z}_2^{(++)}$ is most easily seen in terms of the Majorana basis defined in Eq. (\ref{eq-maj-basis}), which reads
\begin{equation}\label{eq-P++}
P_{++}=(-1)^{ \int \normord{i \psi_1(x) \psi_3(x)} dx }=\pm1\ .
\end{equation}
Similarly, the above three parities and the total momentum $K_\text{tot}$ are not independent, as Eq. (\ref{eq-total-parity}) implies. Therefore, a complete set of independent symmetry charges is $K_\text{tot}$, $N_\uparrow$ and $P_{++}$.

In the bosonized representation, the model now has two nonlinear sine or cosine terms given by $J_\uparrow$ and $J_\downarrow$:
\begin{equation}
J_\uparrow e^{2i\phi_\uparrow}+J_\downarrow e^{2i\phi_\downarrow}+h.c. 
\end{equation}
Therefore, the model is more ``nonlinear" compared to the cases in Secs. \ref{subsec-U1} and \ref{subsec-U1Z2}.

As shown in Fig. \ref{fig-Z2Z2Z2}, the LSS (Fig. \ref{fig-Z2Z2Z2}(c)) in a fixed symmetry sector $(K_\text{tot}=\frac{27}{2},(-1)^{N_\uparrow}=+1,P_{++}=+1)$ (the red part of DOS in Fig. \ref{fig-Z2Z2Z2}(a)) in this case becomes (GOE) Wigner-Dyson statistics. Therefore, we conclude the model in this case is quantum chaotic. The linear ramp at small $\delta_E$ indicates it resembles a GOE distribution (with $m=1$ in Eq. (\ref{eq-WD})). Indeed, the Hamiltonian of the model in this case is a real matrix in the momentum space, as protected by a an anti-unitary PT symmetry, where $P$ is the spatial inversion and $T$ is the spinless time-reversal. Intriguingly, the spectral weights of the model in this case (Fig. \ref{fig-Z2Z2Z2}(b)) is close to that of the free-boson case, despite being a quantum chaotic model.

\subsection{The case with $\mathbb{Z}_{2}^{(\uparrow)}\times\mathbb{Z}_{2}^{(\downarrow)}$ symmetry}\label{subsec-Z2Z2}

The global symmetry of the model can be further lowered down to only $\mathbb{Z}_{2}^{(\uparrow)}\times\mathbb{Z}_{2}^{(\downarrow)}$ if the parameters satisfy
\begin{equation}\label{eq-Z2Z2-condition}
J_s\neq0\ ,\ \ \Delta=0\ , \ \ M_{ss}\neq0\ .\ \ M_{\uparrow\downarrow}=0\ ,\ \  U\neq0\ .
\end{equation}
Compared to the case in Eq. (\ref{eq-Z2Z2Z2-condition}), here the presence of nonzero $M_{\uparrow\uparrow}$ or $M_{\downarrow\downarrow}$ breaks the conservation of the parity charge in Eq. (\ref{eq-P++}). Therefore, the independent symmetry charges in this case are $K_\text{tot}$ and $(-1)^{N_\uparrow}$.

\begin{figure}[tbp]
\begin{center}
\includegraphics[width=3.3in]{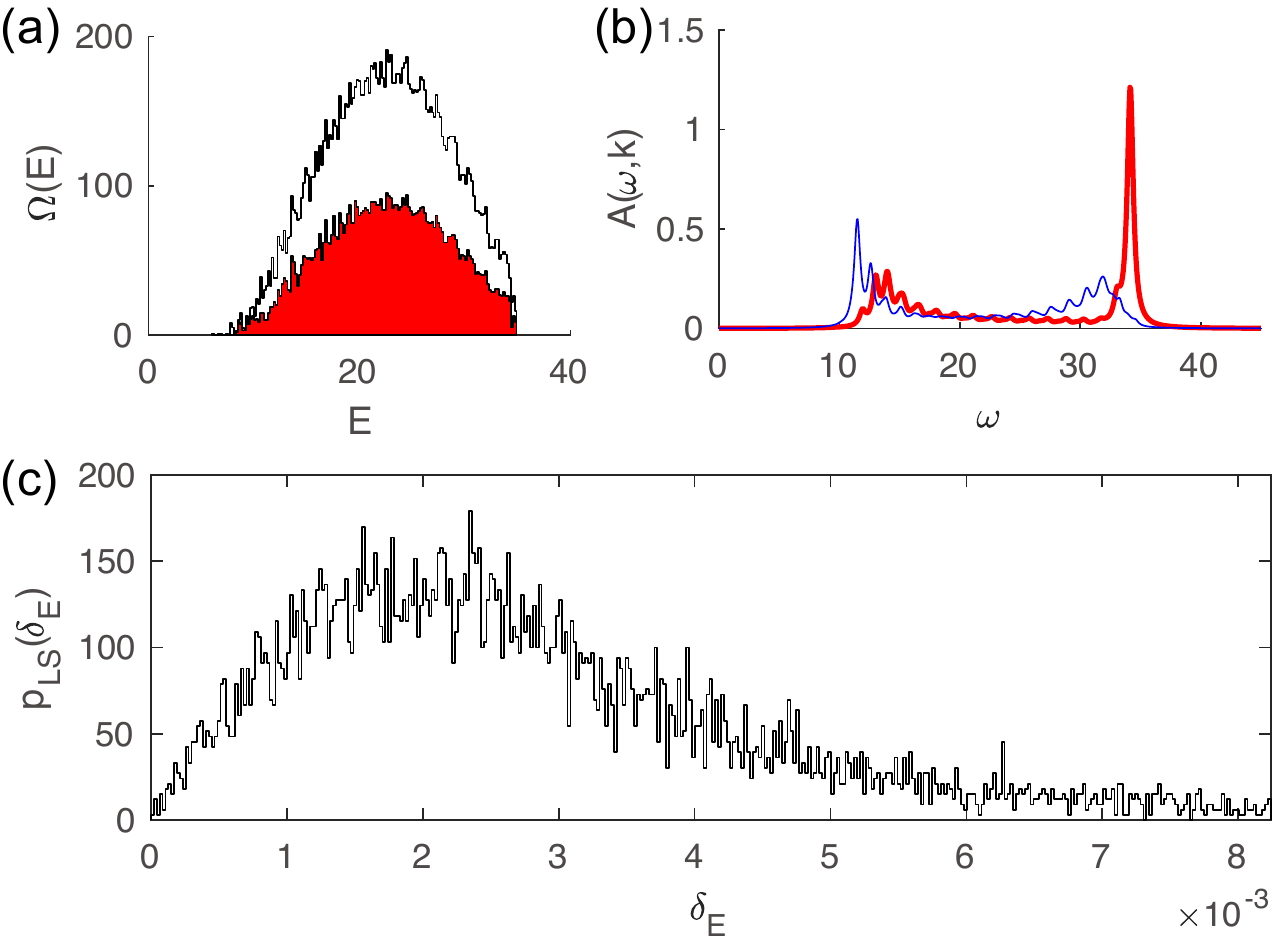}
\end{center}
\caption{ED calculation in the case with $\mathbb{Z}_{2}^{(\uparrow)}\times\mathbb{Z}_{2}^{(\downarrow)}$ symmetry. The parameters are set as $U=1.2\pi$,  $v_\uparrow=1.8$, $v_\downarrow=1.55$, $M_{\uparrow\uparrow}=1$, $M_{\downarrow\downarrow}=1$, $M_{\uparrow\downarrow}=0$, $J_\uparrow=0.5$, $J_\downarrow=0.45$, and $\Delta=0$. (a) The DOS of the $K_\text{tot}=\frac{27}{2}$ sector (unfilled line) and the subsector with $(-1)^{N_\uparrow}=+1$ and thus $(-1)^{N_\downarrow}=(-1)^{N_\uparrow+2K_\text{tot}}=-1$ (the line filled with red). (b) The spectral weights $A_s(\omega,k)$ of spin up (red thick line) and spin down (blue thin line) fermions at $k=\frac{27}{2}$. (c) The LSS of the symmetry sector of $(K_\text{tot}=\frac{27}{2},(-1)^{N_\uparrow}=+1)$ (DOS given as the red part in (a)).
}
\label{fig-Z2Z2}
\end{figure}

The ED results for this case is shown in Fig. \ref{fig-Z2Z2}. Fig. \ref{fig-Z2Z2}(a) shows the DOS of the total sector of momentum $K_\text{tot}=\frac{27}{2}$ and the symmetry subsector with $K_\text{tot}=\frac{27}{2},(-1)^{N_\uparrow}=+1$. The LSS of this symmetry subsector is given in Fig. \ref{fig-Z2Z2}(c), which shows a GOE (linear at small $\delta_E$) Wigner-Dyson shape. Therefore, similar to the case in Sec. \ref{subsec-Z2Z2Z2}, the model here with $\mathbb{Z}_{2}^{(\uparrow)}\times\mathbb{Z}_{2}^{(\downarrow)}$ symmetry is also quantum chaotic. The GOE distribution is also due to the presence of a PT symmetry, which restricts the Hamiltonian in the momentum space to be real. The spectral weights show clear deviations from that in the free-boson case.

\subsection{The case with $\mathbb{Z}_{2}$ symmetry}

In the last case, if we do not impose any constraints on the parameters, the model only has a global fermion parity $\mathbb{Z}_2$ symmetry, the symmetry charge of which is $(-1)^N=(-1)^{N_\uparrow+N_\downarrow}=(-1)^{2K_{\text{tot}}}$ (Eq. (\ref{eq-total-parity})). Therefore, the only independent conserved charge is $K_{\text{tot}}$.

\begin{figure}[tbp]
\begin{center}
\includegraphics[width=3.3in]{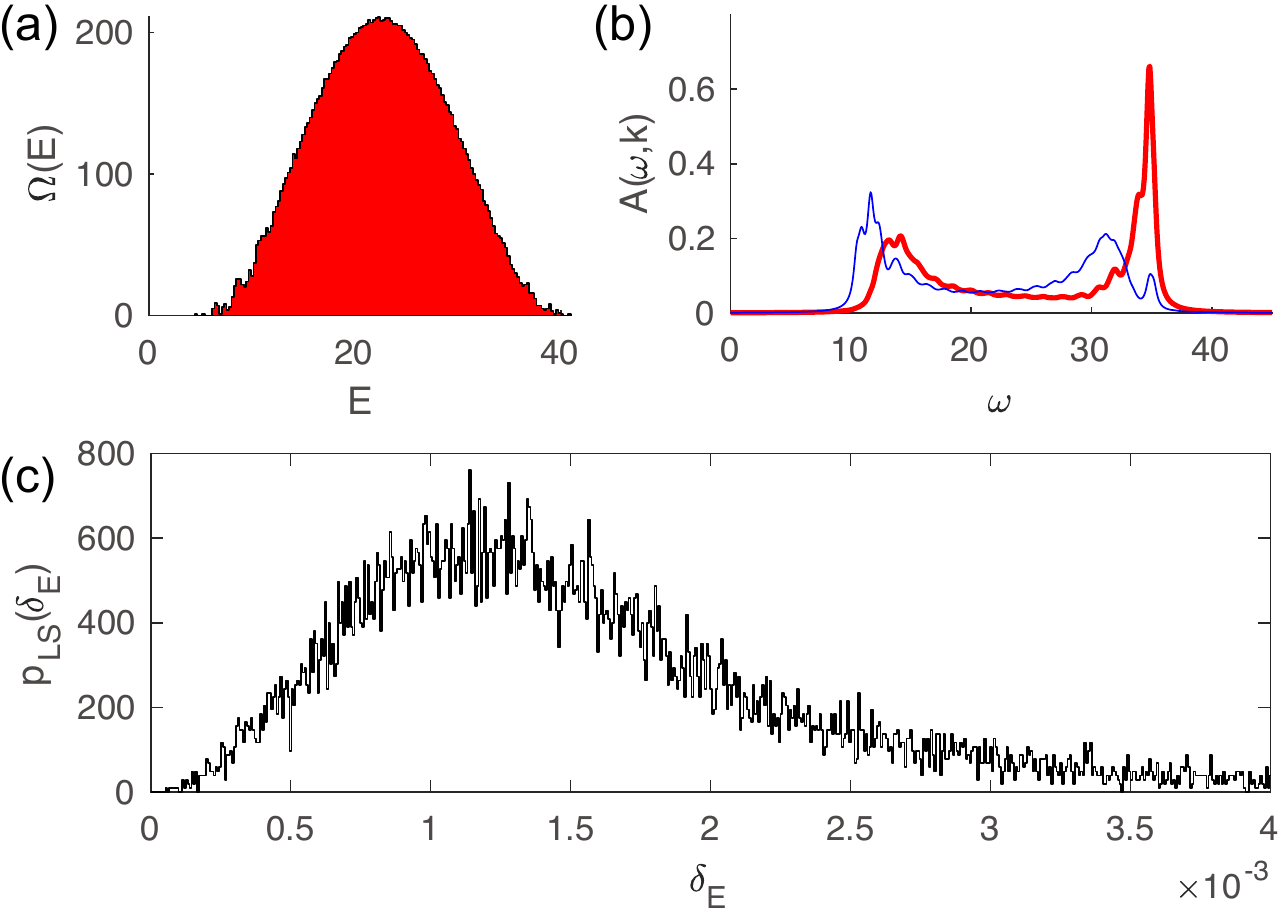}
\end{center}
\caption{ED calculation in the most generic case with only $\mathbb{Z}_{2}$ symmetry. The parameters are chosen as $U=1.2\pi$,  $v_\uparrow=1.8$, $v_\downarrow=1.55$, $M_{\uparrow\uparrow}=1$, $M_{\downarrow\downarrow}=1$, $M_{\uparrow\downarrow}=1+0.5i$, $J_\uparrow=0.5$, $J_\downarrow=0.45$, and $\Delta=0.5+0.3i$. (a) The DOS of the $K_\text{tot}=\frac{27}{2}$ sector (filled with red). (b) The spectral weights $A_s(\omega,k)$ of spin up (red thick line) and spin down (blue thin line) fermions at $k=\frac{27}{2}$. (c) The LSS of the symmetry sector of $K_\text{tot}=\frac{27}{2}$.
}
\label{fig-Z2}
\end{figure}

Fig. \ref{fig-Z2} shows a ED calculation for parameters (see the caption) in this generic case. In the symmetry sector of total momentum $K_\text{tot}=\frac{27}{2}$, the DOS distribution (Fig. \ref{fig-Z2}(a)) is much smoother than all the higher symmetry cases we discussed earlier. The fermion spectral weights in Fig. \ref{fig-Z2}(b) also shows less discretized peaks., indicating a higher randomness in the energy spectrum. Fig. \ref{fig-Z2}(c) shows the LSS of the $K_\text{tot}=\frac{27}{2}$ sector, which is quadratic at small $\delta_E$, and thus resembles the GUE Wigner-Dyson distribution (namely, $m=2$ in Eq. (\ref{eq-WD})). This is because, with the parameters $M_{\uparrow\downarrow}$ and $\Delta$ being generically complex, the model does not have a PT symmetry or other anti-unitary symmetry, and the Hamiltonian is generically in the complex class. Therefore, one expects a GUE LSS in each symmetry charge sector. The model with only a $\mathbb{Z}_2$ symmetry is therefore many-body quantum chaotic.

\section{The effect of nonlinear dispersion}\label{sec-nonlinear}

It is worthwhile to compare the numerical results of our model in the above cases in Sec. \ref{sec-generic} with the interacting model with a nonlinear dispersion added. To be explicit, we start with the free-boson solvable point, namely, the Hamiltonian in Eq. (\ref{eq:H}) with parameters satisfying Eq. (\ref{eq-boson-condition}) (which has U(1)$^{(\uparrow)}\times$U(1)$^{(\downarrow)}$ symmetry), and add a cubic dispersion term
\begin{equation}
\mathcal{H}_{nl}=i\lambda\sum_{s}c^\dag_s \partial_x^3 c_s\ ,
\end{equation}
where $\lambda$ is the the coupling strength. This yields a free-fermion dispersion $\omega_s(k)=v_sk+\lambda k^3$. Such nonlinear terms are irrelevant, so one expects it not to affect the low energy physics. This term does not affect the global symmetry of the model. However, this nonlinear term will break the quantum integrability of the model (at energy scales where this term cannot be ignored).

\begin{figure}[tbp]
\begin{center}
\includegraphics[width=3.3in]{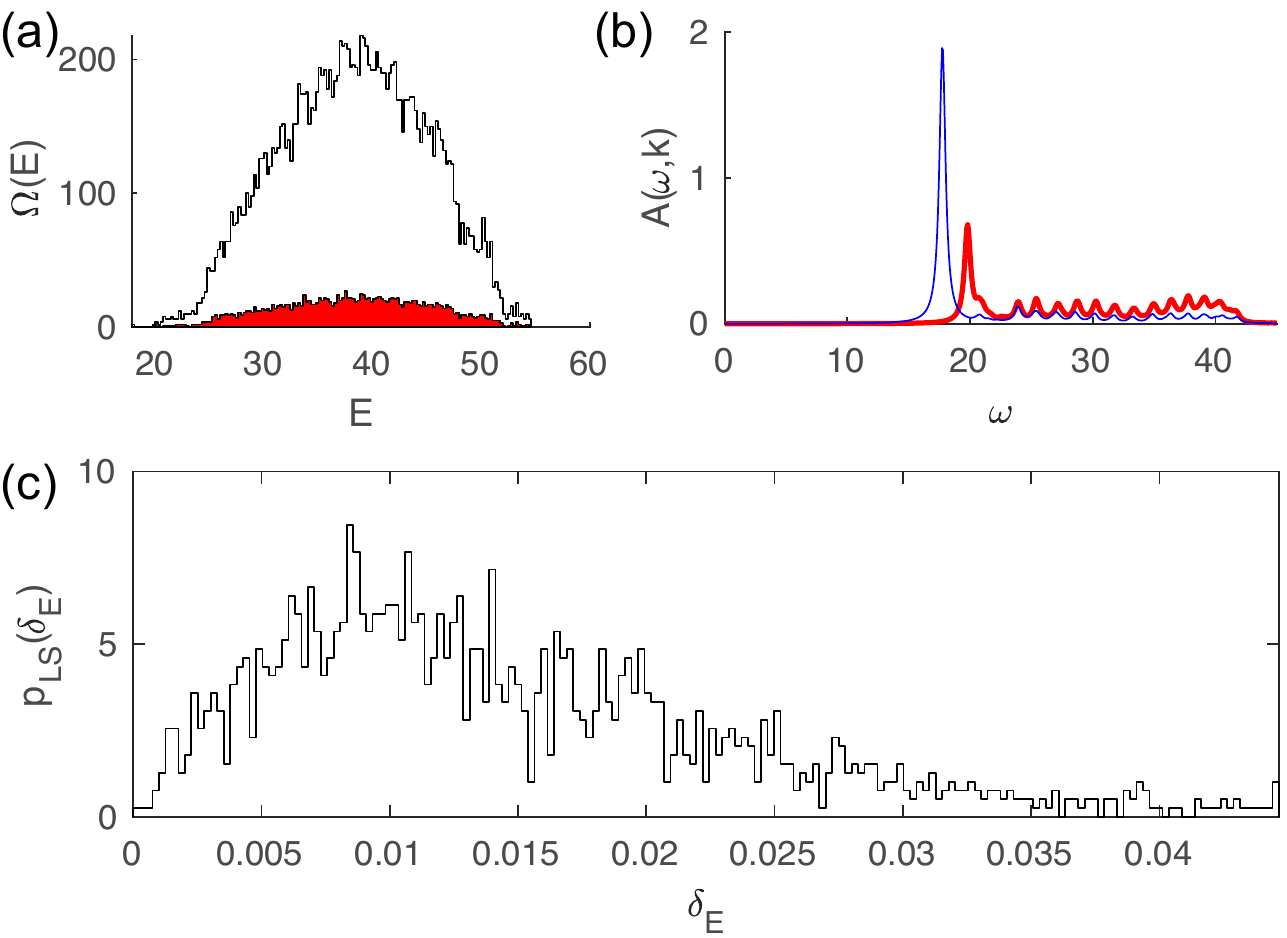}
\end{center}
\caption{ED calculation of the model with a cubic nonlinear dispersion term in the U(1)$^{(\uparrow)}\times$U(1)$^{(\downarrow)}$ symmetry case. The parameters are chosen as $\lambda=20$, $U=1.6\pi$,  $v_\uparrow=2$, $v_\downarrow=1.5$, $M_{\uparrow\uparrow}=0$, $M_{\downarrow\downarrow}=0$, $M_{\uparrow\downarrow}=0$, $J_\uparrow=0$, $J_\downarrow=0$, and $\Delta=0$. (a) The DOS of the $K_\text{tot}=\frac{27}{2}$ sector (unfilled line) and its subsector with $N_\uparrow=1,N_\downarrow=0$ (line filled with red). (b) The spectral weights $A_s(\omega,k)$ of spin up (red thick line) and spin down (blue thin line) fermions at $k=\frac{27}{2}$. (c) The LSS of the symmetry sector of quantum numbers $(K_\text{tot}=\frac{27}{2},N_\uparrow=1,N_\downarrow=0)$.
}
\label{fig-non}
\end{figure}

Fig. \ref{fig-non} shows the ED results for such a model with a large nonlinear dispersion $\lambda=20$ (the other parameters given in the caption). Compared with the linear dispersion case in Fig. \ref{fig-U1U1}, the nonlinear term makes the DOS much smoother (Fig. \ref{fig-non}(a)), distorts the fermion spectral weights (Fig. \ref{fig-non}(b)), and changes the LSS in each $(K_\text{tot},N_\uparrow,N_\downarrow)$ symmetry charge sector into a GOE Wigner-Dyson distribution. Therefore, the model shows quantum chaos due to the nonlinear dispersion term.

An intriguing case which can be compared with the current case is the U(1) symmetric case we studied in Sec. \ref{subsec-U1}, which has a generic nonzero $M_{ss'}$ matrix. Diagonalizing the free fermion part of the U(1) symmetric case Hamiltonian (i.e., Eq. (\ref{eq-free-h})) yields a nonlinear fermion dispersion $\omega_\pm(k)=\frac{v_\uparrow+v_\downarrow}{2}k+\frac{M_{\uparrow\uparrow}+M_{\downarrow\downarrow}}{2}\pm\sqrt{[\frac{(v_\uparrow-v_\downarrow)k+(M_{\uparrow\uparrow}-M_{\downarrow\downarrow})}{2}]^2+M_{\uparrow\downarrow}^2}$ in the eigen-fermion basis. Therefore, it is also an interacting fermion problem with nonlinear dispersions. However, in the U(1) symmetric case, each of the symmetry charge sector shows Poisson LSS, which is drastically different from the nonlinear model in this section. Therefore, the U(1) symmetric case in Sec. \ref{subsec-U1} is a special nonlinear dispersion model with hidden conserved quantities.

\section{Discussion}\label{sec:discussion}

\begin{figure}[tbp]
\begin{center}
\includegraphics[width=3.3in]{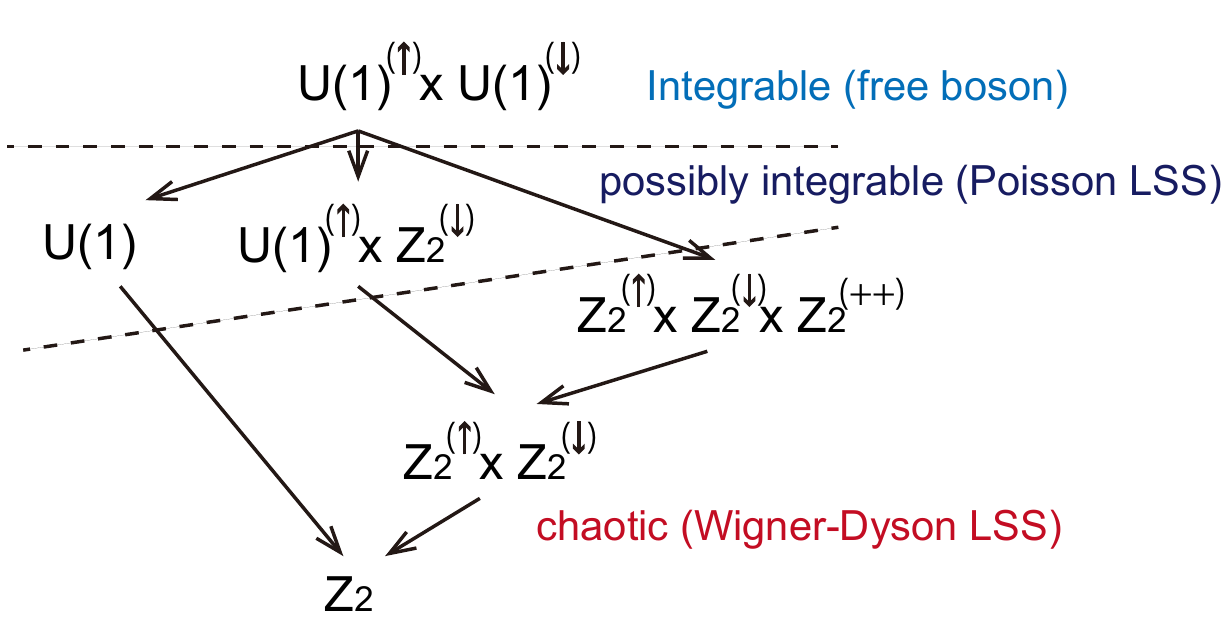}
\end{center}
\caption{Summary of the quantum integrability/chaos transitions of the interacting chiral fermion model in Eq. (\ref{eq:H}) with respect to the global symmetries.
}
\label{fig-sym}
\end{figure}

We have demonstrated that a simple interacting model of two-flavors of chiral fermions shows a rich transition of quantum integrability/chaos with respect to the global symmetries of the model. In the model, we have ignored all the irrelevant terms (except for Sec. \ref{sec-nonlinear}), the effects of which will be suppressed at low energies. Starting from the solvable Luttinger liquid point which yields free chiral bosons with linear dispersions, the lowering of global symmetries leads to a transition to possibly integrable (Poisson LSS in each symmetry sector) energy spectrum and then to quantum chaotic (Wigner-Dyson LSS in each symmetry sector) energy spectrum. Fig. \ref{fig-sym} summarizes this integrable to chaotic transition process versus the global symmetries (the translation symmetry is not listed, which is always there). In particular, the Poisson LSS (possibly integrable) regime indicates there exist hidden (quasi)local conserved quantities, and it would be interesting and useful to explore what they are. It would also be helpful to explore the transition behavior from Poisson to Wigner-Dyson LSS \cite{vir2021} due to symmetry breaking. Numerically, one possible method is to detect such conserved quantities from their eigenstate reduced density matrices (entanglement Hamiltonians) \cite{lian2022conserv}.

One future question is how the integrable or chaotic energy spectra affect the low-energy quantum dynamics of the excitations in such 1D chiral systems, which may be realized as the edge states of 2D topological phases. For the cases with Poisson LSS in each global symmetry sector, the hidden conserved quantities could protect (fully or partially) the quantum coherence of the edge states in certain ways, which may be detectable in edge state interferometer experiments, such as the Fabry-P\'erot interferometer geometry \cite{Nakamura2020} and tunnel junctions \cite{Lian2016,Lian2018a,Fu2009,Akhmerov2009}. In particular, it is possible to control the global symmetries of the chiral edge states experimentally to examine the differences in quantum dynamics with respect to symmetries (Fig. \ref{fig-sym}). For instance, by adding superconducting proximity, one may reduce the symmetry of the model from U(1) to $\mathbb{Z}_2$. Besides, Ref. \cite{hu2021integrability} shows that the model in the U(1)$^{(\uparrow)}\times\mathbb{Z}_2^{(\downarrow)}$ symmetry case is equivalent to the interacting chiral edge states of the 4/3 FQH state and a class of other bilayer FQH states.

The present model can be further generalized into fractionalized anyonic models. such as the FQH edge states \cite{hu2021integrability,naud2000} and Wess-Zumino-Witten (WZW) models \cite{wess1971,witten1983,novikov1982,witten1984,hu2021chiral}. An example of large number of interacting WZW models is studied in Ref. \cite{hu2021chiral}. Moreover, the effect of spatial disorders on the quantum integrability are yet to be investigated, which, however, will break the translational symmetry and makes the ED numerical calculations extremely difficult. Thus, new methods are desired for studying such systems.

\begin{acknowledgments}
\emph{Acknowledgments.} The author is honored to contribute to the Chen-Ning Yang Centenary Festschrift. This work is supported by the Alfred P. Sloan Foundation, and NSF through the Princeton University’s Materials Research Science and Engineering Center DMR-2011750.
\end{acknowledgments} 

\bibliographystyle{unsrt}
\bibliography{Lutt_ref}

\end{document}